\documentclass{emulateapj}

\usepackage{color}
\newcommand{\arcsecs}{\hbox{$^{\prime\prime}$}}
\def\etal{{\it et al.}}

\begin{document}

\title{Ellerman Bombs - Evidence for Magnetic Reconnection \\ in the Lower Solar Atmosphere}
\author{C.J. {Nelson}\altaffilmark{1,2}, S. Shelyag\altaffilmark{3}, M. Mathioudakis\altaffilmark{4}, J.G. Doyle\altaffilmark{1}, M.S. Madjarska\altaffilmark{1}, H. Uitenbroek\altaffilmark{5}, R. Erd\'elyi\altaffilmark{2}.}

\shortauthors{Nelson \it{et al.}}
\shorttitle{Excitation of Ellerman Bombs}
\altaffiltext{1} {Armagh Observatory, College Hill, Armagh, UK, BT61 9DG.}
\altaffiltext{2} {Solar Physics and Space Plasma Research Centre, University of Sheffield, Hicks Building, Hounsfield Road, Sheffield, UK, S3 7RH.}
\altaffiltext{3} {Monash Centre for Astrophysics, School of Mathematical Sciences, Monash University, Clayton, Victoria, Australia, 3800.}
\altaffiltext{4} {Astrophysical Research Centre, School of Mathematics and Physics, Queen's University, Belfast, UK, BT7 1NN.}
\altaffiltext{5} {National Solar Observatory, Sacramento Peak, P.O. Box 62, Sunpsot, NM, USA, 88349.}

\begin{abstract}
The presence of photospheric magnetic reconnection has long been thought to give rise to short and impulsive events, such as Ellerman bombs (EBs) and \ion{Type}{2} spicules. In this article, we combine high-resolution, high-cadence observations from the {\it Interferometric BIdimensional Spectrometer} (IBIS) and {\it Rapid Oscillations in the Solar Atmosphere} (ROSA) instruments at the Dunn Solar Telescope, National Solar Observatory, New Mexico with co-aligned {\it Atmospheric Imaging Assembly} (SDO/AIA) and {\it Solar Optical Telescope} (Hinode/SOT) data to observe small-scale events situated within an active region. These data are then compared with state-of-the-art numerical simulations of the lower atmosphere made using the MURaM code. It is found that brightenings, in both the observations and the simulations, of the wings of the H$\alpha$ line profile, interpreted as EBs, are often spatially correlated with increases in the intensity of the \ion{Fe}{1} $6302.5$ \AA\ line core. Bi-polar regions inferred from Hinode/SOT magnetic field data show evidence of flux cancellation associated, co-spatially, with these EBs, suggesting magnetic reconnection could be a driver of these high-energy events. Through the analysis of similar events in the simulated lower atmosphere, we are able to infer that line profiles analogous to the observations occur co-spatially with regions of strong opposite polarity magnetic flux. These observed events and their simulated counterparts are interpreted as evidence of photospheric magnetic reconnection at scales observable using current observational instrumentation.
\end{abstract}
\keywords{Ellerman Bombs - Magnetic Reconnection - Numerical Simulations}

\section{INTRODUCTION}

Ellerman bombs  (EBs) were first observed by \citet{Ellerman17} and are small-scale ($1$\arcsecs\ or less), short-lived ($2$-$15$ minutes), impulsive events detected in the lower solar atmosphere (see, {\it e.g.}: \citealt{Zachariadis87}; \citealt{Georgoulis02}; \citealt{Watanabe11}; \citealt{Nelson13}). It has been widely suggested that a link exists between EBs and photospheric vertical magnetic fields; \citet{Pariat04} presented co-aligned magnetograms and observations showing the formation of EBs in the plage region trailing an emerging active region (AR). More recently, \citet{Nelson13} found that strong H$\alpha$ line wing enhancements, identified as small-scale EBs, almost ubiquitously surrounded a complex penumbral structure in an emerging AR and are linked to strong magnetic fields and G-band magnetic bright points (MBPs). Due to the link between EBs and strong photospheric magnetic fields, it has often been asserted that EBs arise as a result of magnetic reconnection in the photosphere. \citet{Georgoulis02} suggested three cartoon topologies which could excite magnetic reconnection in the photosphere, including the partial sinking of a flux tube due to inter-granular down-flows, flux loops emerging in a {\it serpentine} manner, and complex uni-polar magnetic fields. Within any AR, many examples of cancellation can be observed by magnetogram data implying the rapid change of magnetic field configuration within the photosphere, potentially consistent with the topologies suggested by \citet{Georgoulis02}. In this article, we address any potential links between EBs and cancellation events around the lead sunspot of a stable AR. 

In a recent review, \citet{Rutten13} suggested that certain brightenings in the wings of the H$\alpha$ line often regarded as EBs could, in fact, be formed due to the influence of high magnetic field concentrations on the H$\alpha$ profile. It was suggested that classical EBs, where energy release leads to increased intensity in the H$\alpha$ line wings, may often be confused with pseudo-EBs, where the line wings of the H$\alpha$ profile outline strong magnetic fields in the lower photosphere. The definitions of these two forms of brightenings advances the work of \citet{Watanabe11}, who suggested that an event must show `flaring', rapid and small scale topological variations associated with high-energy, to be classified as an EB. Due to the sparseness of high-resolution magnetogram data, however, it has so far proved difficult to identify whether certain topologies lead to the classical EB form, {\it i.e.}, whether only bi-polar regions lead to `flaring' events.

\begin{figure*}
\epsscale{1.25}
\center
\plotone{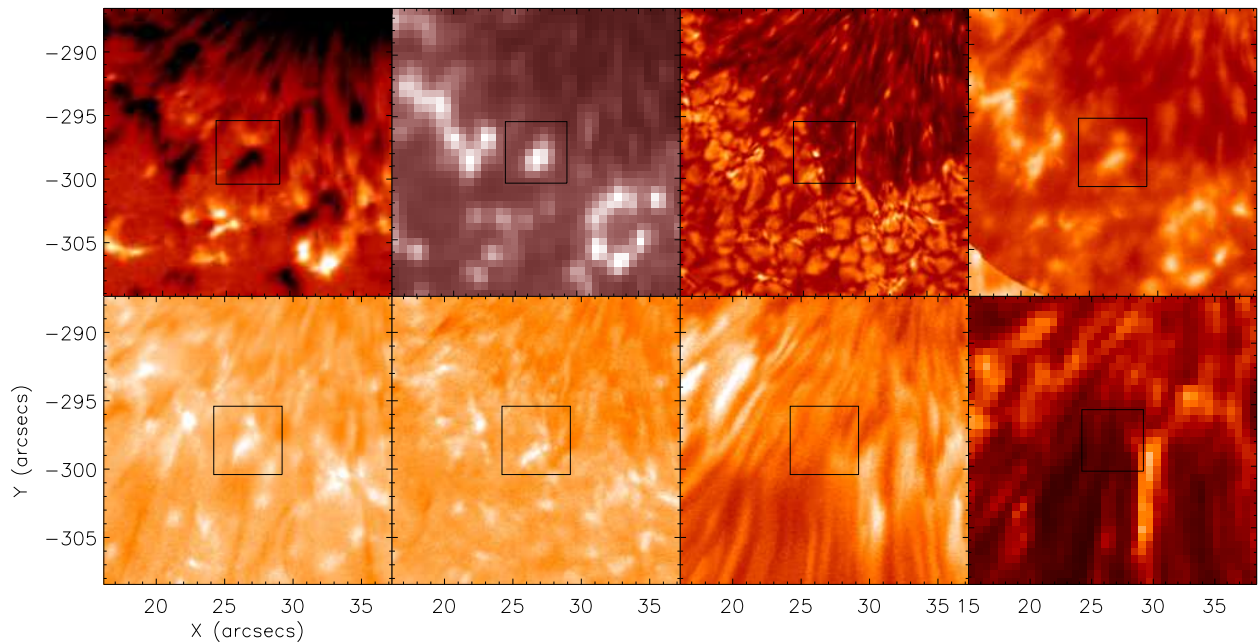}
\caption{(Top row - left to right) Coaligned Hinode/SOT Stokes V/I, SDO/AIA $1700$ \AA, ROSA G-band, and IBIS \ion{Fe}{1} line core. (Bottom row - left to right) IBIS H$\alpha$ blue and red wing images (approximately $\pm 0.75$ \AA), IBIS H$\alpha$ line core, and SDO/AIA $304$ \AA. A box including the event analysed in Fig.~\ref{fig3}, Fig~\ref{fig6}, and Fig.~\ref{fig7} is overplotted. This FOV is taken at the temporally closest image for each wavelength to 14:56:35 UT.}
\label{fig1}
\end{figure*}

Strong observational evidence supporting a photospheric magnetic reconnection model has, so far, proved difficult to establish. The spatial and temporal resolutions of modern magnetogram data are low compared to hypothesised reconnection events (such as EBs), as well as events within the simulated photosphere (see, {\it e.g.}, \citealt{shelyag_lines}), meaning unequivocal inferences about observed magnetic topologies and evolution are rare. In recent years, however, the possible importance of reconnection in the photosphere has been highlighted by the suggestion that such high-energy events could be driving mass into the chromosphere through the excitation of, for example, \ion{Type}{2} spicules \citep{dePontieu07}, as well as, potentially, providing energy for heating. This, in turn, has highlighted the need for an analysis of the lower solar atmosphere in order to infer if any observational signatures of reconnection exist.

Numerical simulations of the lower solar atmosphere have been used to investigate the link between brightening events in the photosphere and bi-polar magnetic fields. \citet{Isobe07}, using the Coordinate Astronomical Numerical Softwares (CANS) code, suggested that an emerging flux loop could form a wide range of reconnection events by propagating upwards from the photosphere, where EBs can be excited, into the corona, where X-ray jets are formed. A $3$-D version of the CANs code was used by \citet{Archontis09} who found that $U$- and $V$-shaped magnetic topologies (similar to those suggested by \citealt{Georgoulis02}) were co-spatial to an increase in temperature in the lower solar atmosphere. These temperature enhancements were hypothesised to be comparable to EB events in the H$\alpha$ line wings, however, line profiles were not simulated. Using a semi-empirical method, \citet{Fang06} found that increases in the VALC temperature in the upper-photosphere and lower-chromosphere lead to the observed intensity enhancements in the wings of the H$\alpha$ and \ion{Ca}{2} $\lambda{8542}$ lines, supporting the conclusions of \citet{Archontis09}.

Recently, \citet{danilovicphd} presented an extensive study of simulated magnetic flux cancellation in the solar photosphere using the low-photospheric absorption lines: \ion{Fe}{1} $6302.5$ \AA, 
\ion{Fe}{2} $5197.58$ \AA, and \ion{Fe}{2} $4923.92$ \AA. One magnetic reconnection event within MPS/University of Chicago Radiative MHD (MURaM) simulations was studied, showing increases in temperature at the inversion line between two opposite polarity regions. Abnormal Stokes V profiles (where, for the \ion{Fe}{1} $6302.5$ \AA\ line, non-rotationally symmetric lobes were returned) as well as signatures of emission in the \ion{Fe}{2} line cores, were found to be co-spatial with such reconnection events. These variations were linked to the process of magnetic field cancellation, formation of current sheets and local Joule heating. Interestingly, \citet{shelyag_lines} found intensity enhancements and splitting of the \ion{Fe}{1} line core co-spatial with magnetic field concentrations in the photosphere, suggesting that signatures of reconnection could be observed in the $6302.5$ \AA\ line profile.

\begin{figure*}
\epsscale{1.2}
\center
\plotone{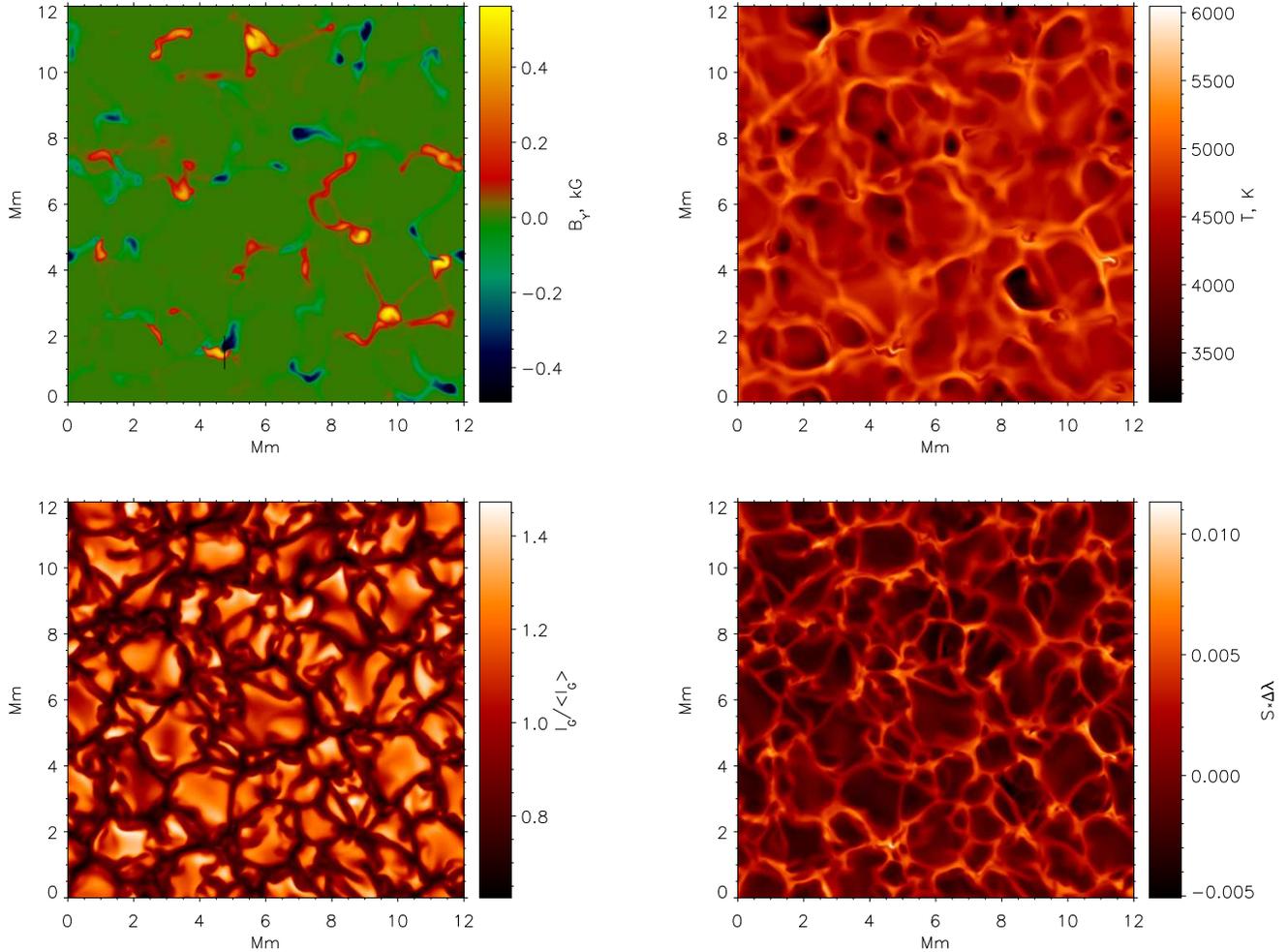}
\caption{(Top row) 
Vertical component of magnetic field (left) and temperature (right) measured at $500~\mathrm{km}$ height above the continuum formation level.
(Bottom row) 
Normalized G-band intensity (left) and \ion{Fe}{1} $6302.5$ {\AA} line strength, $S$, multiplied by the line Doppler-shift, $\Delta\lambda$ (right). 
The reconnection event studied in this paper is marked by a line at (4.75, 1-2) Mm in the top left panel.
}
\label{fig2}
\end{figure*}

In this article, we combine observations and simulations to investigate the link between EBs and the underlying magnetic field. High-resolution, high-cadence multi-wavelength observations of a sunspot and the surrounding plasma, show the link between the observational signatures of EBs in the wings of the H$\alpha$ line and enhancements in the core of the \ion{Fe}{1} $6302.5$ \AA\ profile.  The output of the simulations, made using the MURaM code, is compared with the observational line profiles. The simulated background magnetic fields co-spatial to these line profiles are then studied at high-resolution, allowing for inferences about magnetic  topologies of the field around EB events, that can not be detected using current observational techniques. It is found that co-spatial increases in the \ion{Fe}{1} $6302.5$ \AA\ line core and H$\alpha$ line wings are found over small, bi-polar regions in both the observations and simulations, presenting the strongest indications to date that EBs are formed by photospheric magnetic reconnection. We present this work as follows: In Section $2$, we discuss the observations used in this article; Section $3$ outlines the code; Section $4$ presents our findings before conclusions are stated in Section $5$.

\section{OBSERVATIONS}

The data used in this study were obtained with the {\it Interferometric BIdimensional Spectrometer} (IBIS) and the {\it Rapid Oscillations in the Solar Atmosphere} (ROSA) instruments at the Dunn Solar Telescope in New Mexico USA, during a period of good seeing between 14:50:59 and 15:05:34 on the 30th September 2012.  IBIS ran a $26$-image sequence, repeated at a cadence of $5.5$ seconds during the time series, sampling the atmosphere around AR 11579. This sequence included $17$ H$\alpha$ wavelength points, taken with unequal step sizes ranging between $\pm\ 1.0$ \AA\ from the line centre, and $9$ \ion{Fe}{1} points, again taken in unequal steps, between $6302.4$ \AA\ and $6302.75$ \AA. The pixel-size of the IBIS instrument is $0.098$\arcsecs, giving a spatial-resolution of approximately $0.2$\arcsecs.

\begin{figure*}
{\includegraphics[scale=0.41]{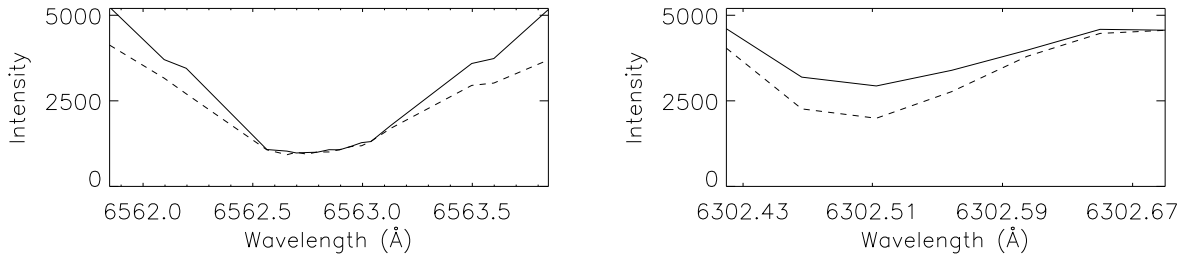}}

\caption{(a) H$\alpha$ line profile for two spatial positions within the box in Fig.~\ref{fig1}. The bold and dashed lines correspond to the EB event and quiet Sun, respectively. (b) Co-spatial \ion{Fe}{1} line profiles.}
\label{fig3}
\end{figure*}

Two ROSA CCDs were employed during this time series, sampling the granulated photosphere, using the G-band filter, and the chromosphere, as observed by \ion{Ca}{2} K (with a band width of $1.2$ \AA\ centred at $3933.7$ \AA). Interpolation time for these two wavelengths was 17 msecs for G-band and 120 msecs for \ion{Ca}{2} K, leading to a total, reconstructed, cadence of 2.112 seconds and 8.448 seconds, respectively. Both wavelengths were observed with diffraction-limited pixel sizes of $0.069$\arcsecs\ giving a spatial resolution of approximately $0.14$\arcsecs. Within the G-band data, numerous groups of MBPs can be observed, situated firmly within the intergranular lanes. Using these MBPs as a proxy for the vertical magnetic field, accurate coalignment to the magnetogram data are possible.

We also made use of observations from two satellites, namely the {\it Solar Dynamics Observatory} and {\it Hinode}. Multi-wavelength analysis is undertaken using the {\it Atmospheric Imaging Assembly} (SDO/AIA). In this study, both the $1600$ \AA\ and $1700$ \AA\ wavelengths are exploited as these have previously been shown to display signatures of EBs (see, {\it e.g.}, \citealt{Qiu00}), however, higher in the atmosphere ({\it e.g.}, $304$ \AA, and hotter), no signal has currently been observed (as is shown in Fig.~\ref{fig1}). We, therefore, limit our analysis to the cooler SDO/AIA lines. Data from the SDO/AIA instrument have physical parameters as follows: pixel sizes of $0.6$\arcsecs, meaning a spatial resolution of $1.2$\arcsecs, and a cadence of $24$ seconds ($12$ seconds for the hotter lines). Magnetic fields are inferred using Stokes V/I data taken by the {\it Solar Optical Telescope} (Hinode/SOT) with a pixel size of $0.155$\arcsecs, a spatial resolution of approximately $0.31$\arcsecs\ and a cadence of around one minute. These data are reduced using the {\it fg\_prep.pro} SSWIDL routine, and then compared to a coaligned slit of spectro-polarimetric data, reduced using the standard {\it sp\_prep.pro} routine. For the quiet Sun, a linear conversion co-efficient (as demonstrated by \citealt{Lites13}) between Stokes V/I and Gauss was found. In this article, we plot Hinode/SOT images with respect to the Stokes V/I parameter.  This is because the full field-of-view (FOV) is not converted to magnetic flux (both the sunspot and penumbra are left in to Stokes V/I form as these create complexities outlined by \citealt{Lites13} and are surplus to the requirements of this research). A zoomed FOV of the region under investigation is shown in Fig.~\ref{fig1}. A box highlighting a typical event (in its infancy) studied in this work is overplotted on the images.

In Fig.~\ref{fig1}, the solar atmosphere is sampled at a number of layers, from the granulated photosphere, observed with the G-band image, to the upper-chromosphere, shown with the SDO/AIA $304$ \AA\ image. A box is overlaid on each wavelength within Fig.~\ref{fig1} to highlight a typical event, as studied in this article. The EB event, identified by an increase in brightness, is seen in the centre of the box in both the H$\alpha$ line wings (approximately $\pm0.75$ \AA\ from the line core), however, it is not observed within the H$\alpha$ line core. The lower photosphere also shows evidence of a co-spatial event, with intensity increases being observed in the SDO/AIA $1700$ \AA\ and ROSA G-band images. Within the G-band data, a small group of MBPs are seen to approach one another in the time preceding this frame before forming this larger, bright event. Links between EB events and these continua have been widely studied in recent years due to the strong dependence which exists between, particularly, G-band and the photospheric vertical magnetic field (see, {\it e.g.}: \citealt{Qiu00}, \citealt{Jess10}, \citealt{Nelson13} for studies of EBs; \citealt{Berger01} for G-band as a magnetic field proxy). Many of these researches have been used as evidence of a firm link between EBs and small-scale fields in support of the results obtained for larger, stronger fields, {\it e.g.}, within a plage region (see, for example, \citealt{Pariat04}) or around a complex penumbra \citep{Zachariadis87}. 

In the upper-photosphere, the \ion{Fe}{1} line core, and lower-chromosphere, \ion{Ca}{2} K images (not shown here),  a co-spatial brightening event is observable; however, no signal of this event is observed in the SDO/AIA $304$ \AA\ images. These observations are typical of EBs and support the hypotheses of \citet{Matsumoto08}, who found evidence that these events were excited in the upper-photosphere. These authors reported flows of opposite direction in the lower-photosphere and lower-chromosphere indicating a high-energy driver situated between the two. This has been supported by simulations (for example, \citealt{Isobe07}; \citealt{Archontis09}) and shall be discussed further with respect to these MURaM simulations.

\section{NUMERICAL SIMULATIONS}

The MURaM  radiative MHD code \citep{voegler1} has been used to carry out simulations of the photosphere for a bi-polar active plage solar region. 
The code
has been thoroughly tested and used for a wide variety of solar and stellar applications \citep[see, {\it e.g.},][]{shelyag_lines, rempel, shelyag_vort, cegla1}. 
It solves the equations of radiative MHD for the large-eddy approximation on a Cartesian grid and employs a 4-th order central difference scheme for calculating
spatial derivatives and a 4-th order Runge-Kutta scheme to advance the solution in time. The numerical scheme is stabilised against numerical instabilities
using hyper-diffusion terms included in each of the MHD equations, with one exception; a small constant diffusion is added for the upper photosphere where hyper-diffusion is suppressed for the magnetic field, as in \citet{voegler1}. The code also uses a short-characteristic, non-grey opacity binning radiation
treatment scheme to account for radiative energy transport in the convection zone and the photosphere. The solar chemical composition is introduced
through the equation of state which takes into account partial ionisation for 11 of the most abundant elements in the solar photosphere.

The numerical box is set to have a horizontal extent of $12 \times 12~\mathrm{Mm^2}$, while the vertical size of the box is $1.4~\mathrm{Mm}$. The
corresponding numerical grid is $480 \times 480$ grid cells in the horizontal directions and $100$ grid cells in the vertical direction, thus the spatial
resolution of the model is $25$ km and $14$ km, respectively. The side boundaries of the numerical domain are periodic.
The top boundary is closed for in- and outflows, while it allows horizontal motions of plasma and magnetic field. The dispersion created by such horizontal flows, and the exponential decrease in density in the upper-layers of the atmosphere, means that the effects of this closed boundary on these simulations is negligible (as was discussed by \citealt{Beeck12}). The bottom boundary is transparent
both for in- and outflows. The $500~\mathrm{nm}$ continuum formation level is located at a height of approximately $800~\mathrm{km}$ above the
bottom boundary.

We start the simulation from a statistically stabilised and well-developed snapshot of non-magnetic photospheric convection. 
In a manner similar to that described by both \citet{khomenko1} and \citet{danilovic1},
a bi-polar, checkerboard magnetic field structure with unsigned magnetic field strength of $200~\mathrm{G}$ is introduced into the domain.
The length of the individual squares of constant magnetic field is chosen to be $2~\mathrm{Mm}$, making a $6 \times 6$ box of opposite magnetic field
regions. This structure is in agreement with the periodic side boundary conditions.

After the magnetic field was introduced into the domain, we let the simulated photosphere evolve for approximately 1.5 hours. During the first
few minutes of the simulation, large parts of the magnetic field were advected into the intergranular lanes, while some magnetic flux was
cancelled due to the opposite-polarity initial configuration \citep[see, {\it e.g.},][]{cameron1}. Magnetic field concentrations of opposite polarity with unsigned 
strength of up to $1.6~\mathrm{kG}$ at the continuum formation level were subsequently formed. Since the turbulent magneto-convection process is effectively
random, these intergranular opposite-polarity magnetic field concentrations move, and sometimes come closer together, reconnect
and cancel out. One typical example of such an event was selected for further investigation.  

Snapshots from our simulations are shown in Fig.~\ref{fig2}. The reconnection region studied in this paper is marked
by a line in the top-left panel.  The vertical component of magnetic field (top-left panel)
and temperature (top-right panel) measured in the photosphere $500~\mathrm{km}$ above the continuum formation level are plotted, and show sharp and localised temperature enhancements in the regions
where opposite polarity magnetic field concentrations come close to each other and reconnect leading to localised flows (as observed by \citealt{Matsumoto08}). Reconnection events within these simulations appear to be of the Sweet-Parker (S-P) type, whereby a current sheet is formed between two regions of opposite polarity field. Although basic implementation of this model provides slow, non-physical reconnection rates, the inclusion of flux pile-ups and strong granular flows around the reconnection point can speed up the process (as has recently been shown by \citealt{Litvinenko07}) to be comparable to physical phenomena, as observed here. Notably, the G-band intensity (bottom-left panel)
does not show any peculiarities in the regions of magnetic reconnection, clearly demonstrating that the significant (compared to the background plasma) 
energy release from magnetic field reconnection occurs higher than the location of the continuum radiation layer. The \ion{Fe}{1} intensity (bottom-right panel) does show an increase in intensity co-spatial to the magnetic reconnection event and, therefore, with the temperature increase.

We performed spectropolarimetric diagnostics of the simulated event with the \ion{Fe}{1} $6302.5$ \AA\ magnetically-sensitive absorption line and the H$\alpha$ line. 
The \ion{Fe}{1} line is known to form in the low photosphere, where, in general, the local thermodynamic equilibrium approximation
holds; however, as reported by \citet{schukina2001}, application of an LTE approximation to neutral iron lines leads to a different temperature sensitivity when compared to
a precise, but computationally difficult, non-LTE approach. Due to expected temperature effects during the reconnection process in the solar photosphere, 
we employed a full non-LTE line profile synthesis code RH
(see \citealt{Uitenbroek01} for information) to perform radiative diagnostics with the \ion{Fe}{1} line.

\begin{figure}
\epsscale{1.2}
\center
\plotone{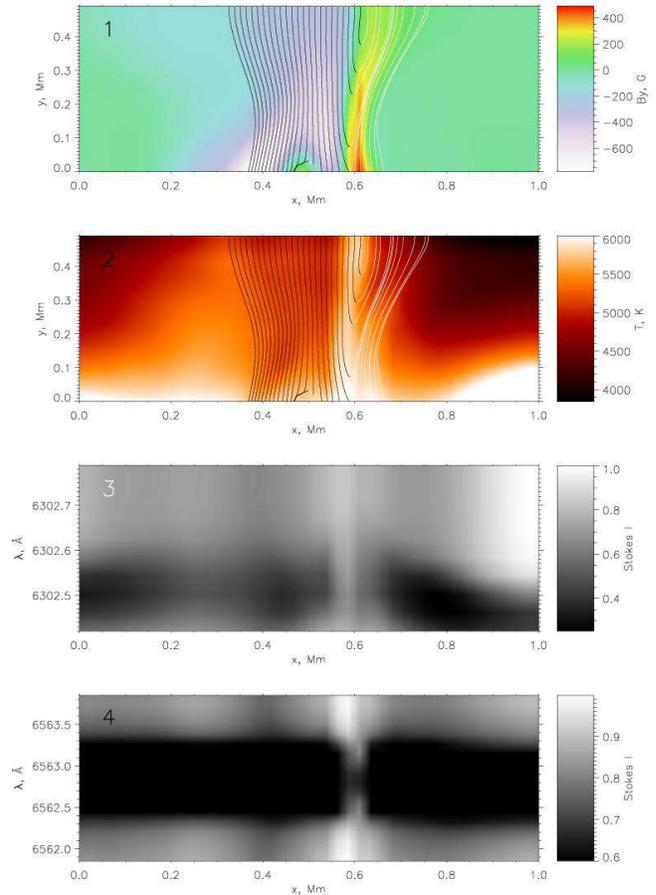}
\caption{A reconnection event in the simulated photosphere. (1) The slice of the simulations box indicated by the slit in Fig.~\ref{fig2} showing the vertical component of magnetic field in the reconnection region with height ($y$-axis where $0$ Mm corresponds to the continuum layer), with the magnetic field lines overplotted. Black lines correspond to the negative (downward) direction of magnetic field and white lines represent positive (upward) direction. The field lines originate in the points where the total magnetic field strength is greater than $150$ G. (2) Temperature image with the field lines overplotted. The reconnection process is clearly seen as a strong temperature enhancement in the region of U-shaped magnetic field lines. (3) \ion{Fe}{1} $6302.5$ \AA\ line profile with a `gap' in the reconnection region ($x \sim 0.6$ Mm) indicating an increase in intensity of the line core. (4) H$\alpha$ line profile shows brightenings in both wings in the reconnection region, which is identified as an EB.}
\label{fig4}
\end{figure}

The H$\alpha$ line is formed over the range of heights starting from the deep photosphere up to the chromosphere or transition region (see, {\it e.g.}, \citealt{Rutten08}, 
\citealt{Leenaarts12}). The wings of the H$\alpha$ line profile are formed in the photosphere and clearly show the presence of various structures, 
from large sunspot umbrae and penumbrae, to smaller rapid blue excursions (see, for example, \citealt{Langangen08}; \citealt{Rouppe09}). What is not clear, however, is whether EBs, observed in the H$\alpha$ wings, are formed in the photosphere as has been 
widely hypothesised (by, {\it e.g.}: \citealt{Georgoulis02,Watanabe11,Nelson13}). In this work, using numerical modelling of a solar atmosphere which includes only the 
photosphere, we aim to model and analyse the behaviour of the H$\alpha$ line wings, thereby proving the formation height of EBs. The same radiative 
diagnostics code RH was used to calculate the H$\alpha$ line profiles.

Since full three-dimensional calculations of the radiative field is still a computationally challenging task, we used a simplified approach of computing the emergent,
wavelength-dependent intensities along the vertical rays in the selected regions of the simulation domain. Then, we degraded the spectral resolution of
the obtained Stokes I profiles in order to directly compare the synthetic images with the observations.

\section{RESULTS \& DISCUSSION}

\begin{figure}
\epsscale{1.2}
\center
\plotone{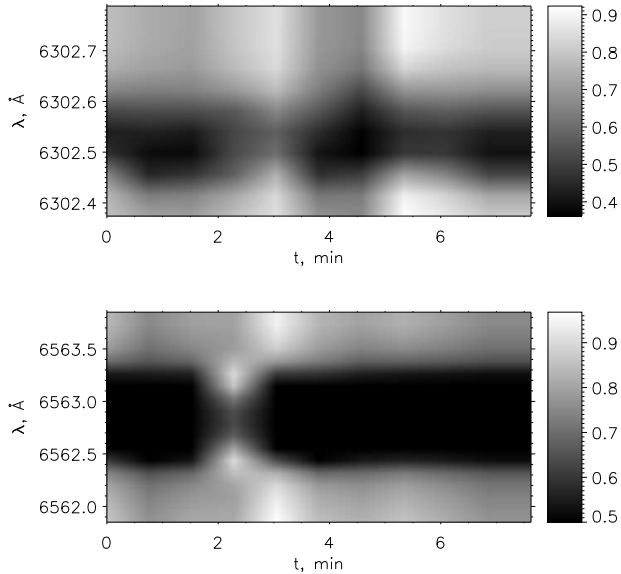}
\caption{Temporal evolution of the simulated EB. (a) \ion{Fe}{1} $6302.5$ \AA\ line profile. (b) H$\alpha$ line. Between 2.5 and 4.5
minutes, the \ion{Fe}{1} line shows a `gap' corresponding to the reconnection event and H$\alpha$ shows line wing increases. Note, there is also a slight
brightening in H$\alpha$ core about one minute before the EB event.}
\label{fig5}
\end{figure}

\begin{figure*}
{\includegraphics[scale=0.5]{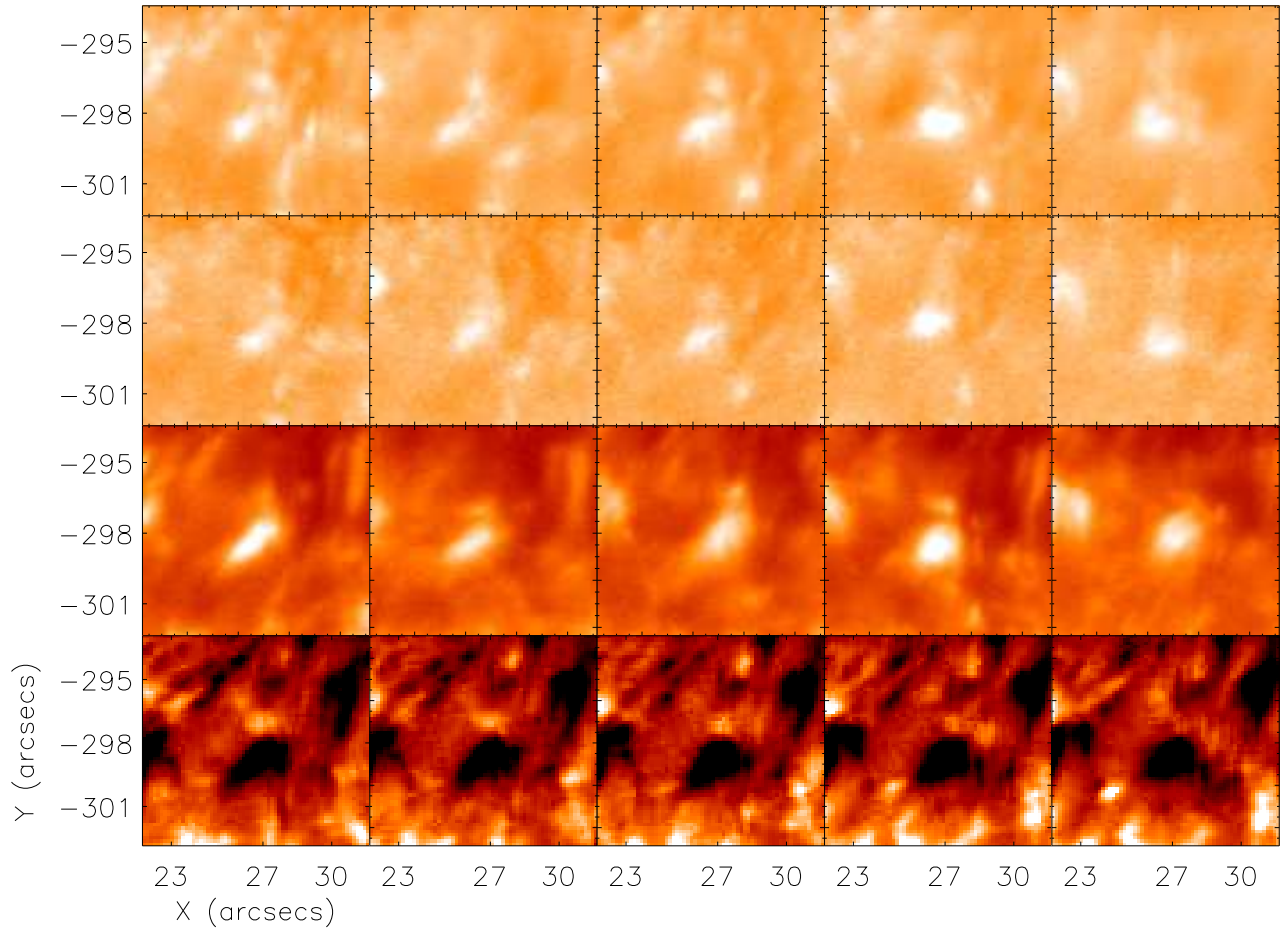}}

\caption{A zoomed FOV of a bi-polar region and a corresponding EB event in the H$\alpha$ blue wing position (top row), H$\alpha$ red wing position (second row), \ion{Fe}{1} line core (third row) and SOT Stokes V/I (bottom row) over time. Each column is the temporally closest frame for each wavelength to 14:52:36, 14:55:38, 14:58:39, 15:01:39, and 15:04:45, respectively. Lightcurves for the full observational period are plotted in Fig.~\ref{fig7}.}
\label{fig6}
\end{figure*}

We identify a number of large EB events within these data, described in Section 2, by analysing H$\alpha$ line profiles for intensity enhancements and assessing for any evidence of `flaring', or an explosive nature, during bright events. Events which show significant increases in line wing intensity, and no signature in the line core, are studied (one representative example of such an event is shown in imaging form within the box in Fig.~\ref{fig1} and line profile form by Fig.~\ref{fig3}). Small, short-lived intensity enhancement events (exhibiting line profiles analogous to EBs), with circular diameters less than $0.5$\arcsecs\ and lifetimes often below $2$ minutes, are common in this dataset (as predicted by \citealt{Nelson13}); however, due to the tendency of line scan data to change quality rapidly, a large-scale statistical study of these short-lived events could provide spurious results. We, therefore, limit our analysis to larger EB events, which show sustained brightening and `flaring' as discussed by \citet{Watanabe11}. It should be noted that no events appear to be brighter than $150$\% of the background intensity of these data (possibly due to effects of seeing) and, hence, $140$\% of the background emission is used as a guide for identification of EBs.

\begin{figure}
\hspace{-20pt}
{\includegraphics[scale=0.38]{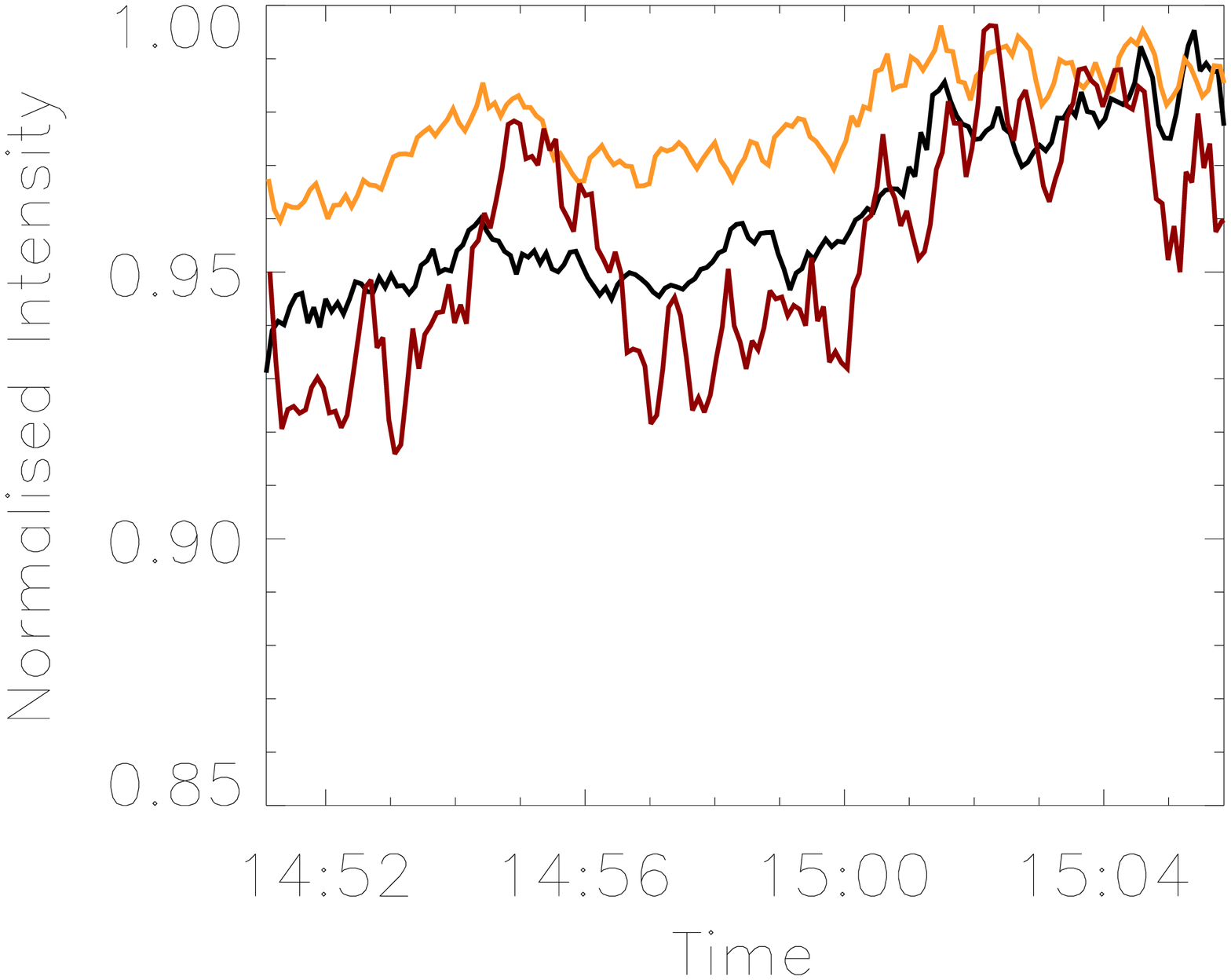}}

\hspace{-20pt}
{\includegraphics[scale=0.38]{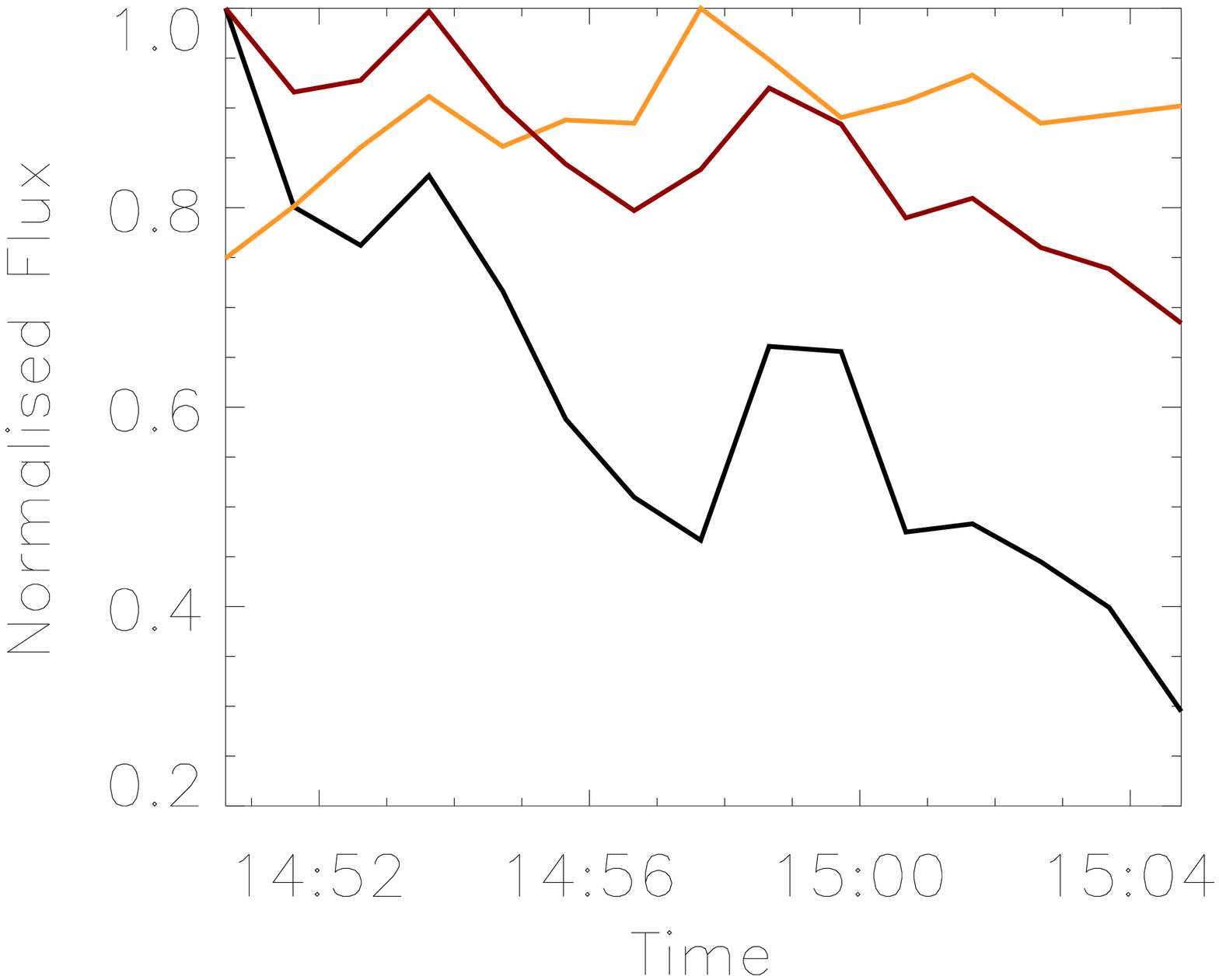}}

\caption{(a) The evolution of IBIS data over time. Black is for the H$\alpha$ blue wing, orange is for the H$\alpha$ red wing and maroon is for the \ion{Fe}{1} line core. (b) The evolution of the magnetic field over time. Black is for negative flux, orange for positive flux and maroon for total flux. A decrease of flux to around $68$ $\%$ of the original strength is observed.}
\label{fig7}
\end{figure}

In Fig.~\ref{fig3}, we plot the H$\alpha$ and \ion{Fe}{1} line profiles for both the EB and a quiet Sun region within the box studied in Fig~\ref{fig1}. The H$\alpha$ line wings show significant intensity enhancements compared to the quiet Sun, whereas, no signature within the line core is observed. This agrees with the definition of EBs put forward by \citet{Ellerman17}, who suggested that these events were ``a very brilliant and very narrow band extending four or five \AA\ on either side of the line, but not crossing it''. Co-spatially, the \ion{Fe}{1} line core shows increases in intensity. This line profile pairing (increases in the H$\alpha$ wings and \ion{Fe}{1} core) is typical of the response of the \ion{Fe}{1} $6302.5$ \AA\ line to the studied EB events in these observations, suggesting a possible link between the formation of the H$\alpha$ and \ion{Fe}{1} lines. We present and analyse one event with these characteristics in more detail in Fig.~\ref{fig6} and Fig.~\ref{fig7}.

Similar, co-spatial behaviour in the H$\alpha$ and \ion{Fe}{1} $6302.5$ \AA\ line profiles is observed in numerical simulations at regions of magnetic field reconnection in the solar photosphere. In Fig.~\ref{fig4}, we plot the magnetic field geometry (panel 1), the temperature structure (panel 2), and the corresponding \ion{Fe}{1} $6302. 5$ \AA\ (panel 3) and H$\alpha$ (panel 4) line profiles for each  vertical cut across the magnetic reconnection region identified in Fig.~\ref{fig2}. The magnetic field lines clearly outline the inversion line between the regions of the opposite magnetic field polarities (the colour of the field lines represents their direction, which is white and black for upward- and downward-directed magnetic field, respectively). The temperature increase, co-spatial with the inversion line, corresponds to the Joule heating and energy release during reconnection \citep{danilovicphd}. The \ion{Fe}{1} line profile shows a prominent `gap' with strongly reduced absorption in the increased temperature region. Co-spatially, the H$\alpha$ line profile indicates a strong increase in intensity in both its wings, similar to an observed EB. These results support the semi-emperical assertions of  \citet{Fang06}, who found that heightened emissions in the H$\alpha$ line wings were most likely a result of increased temperature within a magnetic inversion line. The above follows on from earlier radiative transfer modelling of Hydrogen spectrum in the dMe stars by \citet{Houdebine94}. Also, a (stronger) narrowing of the profile, as in the observations, is simulated. This narrowing coincides with the uppermost part of the reconnection layer.  The narrowing is formed due to the heightened electron density at the flare reconnection site increasing the atmospheric mass load, thus creating a deeper absorption core \citep{Cram79}.

The time-dependent nature of reconnection, the evolution of the magnetic field geometry in the simulations, and the short-lived modification of the H$\alpha$ line profile in observed EBs suggests further investigation into the temporal properties of these events is necessary. Namely, the expansion of the magnetic field concentrations towards the upper layers of the solar atmosphere due to decreasing
gas pressure will cause the reconnection to happen earlier in the upper-simulated photosphere than in the lower regions. Thus, the thermal effect of reconnection will first be seen in the parts of the H$\alpha$ line profile sensitive to the higher solar atmosphere, followed by intensity increases in the H$\alpha$ wings and reduced opacity in the \ion{Fe}{1} $6302.5$ \AA\ line core. This process is demonstrated in Fig.~\ref{fig5}, where the temporal evolution of the \ion{Fe}{1} (top panel) and H$\alpha$ (bottom panel) line profiles during the reconnection event are plotted. It is easy to see that the H$\alpha$ line core narrowing (due to increased electron density) is followed by the \ion{Fe}{1} `gap' and the rise in the H$\alpha$ wings, which occur co-temporally.

It should be noted, that the rise in the \ion{Fe}{1} wing intensity following the EB event in Fig.~\ref{fig5} does not necessarily correspond to the same driver. At around 5 minutes, there is a significant increase in the \ion{Fe}{1} red wing intensity which is simultaneous with a very mild increase in the H$\alpha$ wings, while no strong rise in the \ion{Fe}{1} core intensity is observed. This behaviour can be linked to an increase in the background photospheric continuum intensity and reduced opacity in the photospheric magnetic field concentration, which does not experience reconnection \citep[see, {\it e.g.},][]{shelyag_lines}.

The temporal evolution of bright events in these observations was analyzed by identifying possible EB events within the H$\alpha$ line wings, and then producing a sequence of images. These sequences included the H$\alpha$ wings, the \ion{Fe}{1} line core, and Hinode/SOT Stokes V/I images. One excellent example is plotted in Fig.~\ref{fig6}. The H$\alpha$ line wings (blue and red for the top and second rows, respectively) and the \ion{Fe}{1} line core (third row) show strong brightenings over the bi-polar region (in the Hinode/SOT Stokes V/I data plotted in the bottom row). The EB shows increased brightness and area over the lifetime of the event, which is identified as `flaring' (as discussed by \citealt{Watanabe11}). Rapid topology changes are easily observed through the sequence implying a dynamic nature for this event. We find that this behaviour is common in EBs found over bi-polar regions and could be used as an indicator that these events are classical EBs rather than the, possibly more common, pseudo-EBs (which purely mimic the magnetic field in the photosphere; see, {\it e.g.}, \citealt{Rutten13}). Conducting an analysis of dopplergrams (made using the [blue-red]/[blue+red+2] formula; see \citealt{Madjarska09}), we find a common association between EBs and upward shifts in the H$\alpha$ line, analogous to those observed by \citet{Matsumoto08}, however, due to the small number of data points in the wings, we are unable to estimate accurately the velocities of these events. 

Lightcurves for this event are plotted in Fig.~\ref{fig7}. These lightcurves were made by focusing the image used in Fig.~\ref{fig6} onto the opposite polarity region such that no new, strong flux appears within the FOV in the lifetime of these observations. The total positive flux was then taken by summing all positive values within the box ({\it vice versa} for the negative polarity) for each frame before plotting the normalised total flux over time. In Fig.~\ref{fig7}(a) we plot the smoothed H$\alpha$ line wing (black and maroon for $\pm\ 0.75$ \AA\ respectively) and the \ion{Fe}{1} line core (orange) intensities. In Fig.~\ref{fig7}(b) the negative polarity (black), positive polarity (orange) and total flux (maroon) are shown. Over the period of the observations, there is a significant decrease in the flux of this small bi-polar region, dropping to around $68$ $\%$ of the original flux, and only $30$ $\%$ of the negative polarity flux. This decrease is easily observed through the sequence of images presented in Fig.~\ref{fig6}. Within a $15$ minute period, a decrease of this order due to magnetic reconnection would suggest that any corresponding energy release must be strong, and could produce large temperature increases in the surrounding atmosphere. 
 
The IBIS data (in Fig.~\ref{fig7}a) reaches a maximum intensity for each of these lines at approximately $15:02$, maintaining a peak for around two to four minutes. This peak is identified as `flaring' within this EB, associated with both increased intensity and area. The short-lived peak is more evident within the \ion{Fe}{1} line core, where an eight percent increase in intensity is observed within around one minute. Co-temporally, a rapid decrease in negative (and total) polarity flux in Fig.~\ref{fig7}(b), which continues until the end of these data, is observed. A second, smaller peak is observable at approximately $14:54$ and corresponds with a large drop in both negative polarity and total flux within this FOV. Although this co-temporal brightening and flux cancellation occurs for several examples within this dataset, the cadence of the Hinode/SOT magnetograms is too low to make further inferences about whether this is a correlation or a coincidence; therefore, we shall leave this aside, and note that it would be worthy of future study when sufficiently high-cadence magnetograms become available.

H$\alpha$ wing brightening events co-spatial to uni-polar regions, such as the strong field at (33, -302) in Fig.~\ref{fig1}, were also studied within these data. The average area of brightenings over uni-polar fields appeared to be larger than bi-polar EBs. Smaller, shorter-lived intensity variations were also common implying that rapid, high-energy releases may not be leading to these brightenings and, hence, that the increase in intensity in the H$\alpha$ line wings and \ion{Fe}{1} line core may be due to the strength of the magnetic field (analogous to the pseudo-EBs discussed by \citealt{Rutten13} and the reduced opacity simulated by \citealt{shelyag_lines}, respectively). Within such events, the shape of the intensity enhancement is often different between the imaged H$\alpha$ wings (which is not true of the bi-polar EBs, such as the event presented in Fig.~\ref{fig6}). We find that brightenings occurring over uni-polar regions still show significant intensity increases (often above thresholds currently applied within automated EB tracking, such as $130$ \% or $150$ \% of the background intensity) within the H$\alpha$ line wings, however, the line profiles frequently show excess intensity in one wing over the other.

The event at (33, -302) in Fig.~\ref{fig1}, for example, displays co-spatial intensity increases for short periods during these observations and would, using thresholds alone, be classified as an EB for these times; however, for the majority of these observations, the red wing exhibits traits associated with the network, whereas the blue wing is extremely bright (as can be seen in Fig.~\ref{fig1}). We suggest that comparable temporal and morphological evolutions between the H$\alpha$ line wings are an important factor in EB identification (similar to that shown in Fig.~\ref{fig6}) and that any events which show non-comparable evolutions between the wings are, in fact, pseudo-EBs.

\section{CONCLUDING REMARKS}

In this article, we have used multi-instrument, multi-wavelength observations and numerical simulations to investigate the formation mechanism of EBs, small-scale brightening events in the H$\alpha$ line wings, commonly found to occur in emerging ARs. Numerous possible EB events were found within these data by identifying instances of H$\alpha$ line wing intensity increases. The evolution of these events was then studied in order to identify whether they showed instances of `flaring'.

In order to investigate the link between EBs and magnetic fields, small opposite polarity flux regions, observed to be spatially isolated from other strong fields, were analysed. It was found that rapid changes in both magnetic flux polarity (apparent conversion from positive to negative, or {\it vice versa}) and magnetic flux strength (a total decrease in flux) were present during each of the studied brightening events.
The evolution of a large, `flaring' EB event and the corresponding flux region is shown over the $15$ minute observational period studied in this article, finding a quite remarkable decrease in total flux to approximately $68$ \% of the initial strength. This rapid cancellation of flux, and the associated brightenings in this multi-wavelength analysis, implies that a release of magnetic energy into the surrounding plasma by a small-scale magnetic reconnection event is a more likely reason for the reduction in flux than a sinking of tubes back into the convective layer. Unfortunately, even relatively high-cadence, high-resolution magnetograms, such as those studied here, are unable to detect the difference between these two scenarios. 

A model box was constructed to simulate the solar photosphere using the MURaM code of radiative magnetoconvection. Within the model box, the line profiles from small bi-polar regions, similar to the representative example presented within this article exhibited similar traits to EBs forming over bi-polar regions. Further study of the evolution of these small-scale events over time in the simulations was undertaken, showing temperature increases co-spatially with magnetic inversion lines at bi-polar regions. Interestingly, the relationship noted in the observations between the H$\alpha$ and \ion{Fe}{1} line profiles also appeared in the simulations, which showed increased intensity in the line wings and line core, respectively. The reconnection events analysed in this article appear to be of Sweet-Parker type, however, a more specialised study is required to fully understand the mechanism which is occuring.

The influence of the magnetic field on the \ion{Fe}{1} lines has been discussed in previous researches (see, {\it e.g.}, \citealt{shelyag_lines, danilovicphd}). \citet{shelyag_lines} suggested that regions of strong flux, especially close to magnetic inversion lines, where opposite polarites converge, could lead to a heightened core, as well as splitting (where two intensity minima are observed within the line profile), whereas, \citet{danilovicphd} found that abnormal (non-rotationally symmetric) \ion{Fe}{1} $6302.5$ \AA\ Stokes V profiles were co-spatial with both strong magnetic fields and magnetic reconnection in the photosphere. It has, therefore, been suggested that the formation of the \ion{Fe}{1} $6302.5$ \AA\ line profile is intrinsically influenced by the magnetic field. A combination of these results could  be used to intuitively expect coaligned EB and \ion{Fe}{1} brightenings as were found here; however, as the observations used in this article could not be used as a diagnostic tool for Stokes V and, due to both the spatial and spectral resolutions, did not find evidence of splitting within the \ion{Fe}{1} line core, we suggest that further research be carried out when sufficiently high-resolution data are available.

Overall, the observed and simulated photospheres analysed in this article, exhibit analogous line profiles co-spatial to bi-polar regions, namely brightenings within the H$\alpha$ line wings and an increase in intensity in the \ion{Fe}{1} $6302.5$ \AA\ line core. We suggest that this analysis has presented the clearest evidence to date, that the sub-class of brightening events known as EBs in the H$\alpha$ line wings are formed by magnetic reconnection in the solar photosphere. However, it should be noted, that improved temporal, spatial and spectral resolution when inferring both photospheric imaging lines and magnetic field is required before observations are capable of standing alone in identifying magnetic reconnection. 

Finally, it should be noted, that this article does not suggest that all identified brightening events are formed through magnetic reconnection in the photosphere. Several events situated over uni-polar regions within the observations that form the required H$\alpha$ profiles to be classified as EBs (including significant brightness enhancements in the line wings), do not possess any evidence of `flaring', which appears to be essential in identifying the more dynamic, higher-energy EBs. We suggest that to make any inferences on future datasets, the authors must have high-resolution, high-cadence data in a number of wavelengths, as presented here. Further research must be undertaken before conclusions about the wider range of brightening events in the H$\alpha$ line wings can be drawn.

\acknowledgements

Research at the Armagh Observatory is grant-aided by the N. Ireland Dept. of Culture, Arts and Leisure. Sergiy Shelyag’s research is supported by Australian Research Council Future Fellowship and with the assistance of resources provided at the NCI National Facility systems at the Australian National University through Astronomy Australia Limited, and at the Multi-modal Australian ScienceS Imaging and Visualisation Environment (MASSIVE) (www.massive.org.au). Sergiy Shelyag also thanks the Centre for Astrophysics \&
Supercomputing of Swinburne University of Technology (Australia) for the computational resources provided. We thank the NSO for their hospitality and in particular Doug Gilliam for his excellent help during the observations. Data reduction was accomplished with help from Kevin Reardon and Peter Keys. We would also like to thank Friedrich W\"oger for his image
reconstruction code and Dave Jess for his destretching routine. We thank the UK Science and Technology Facilities Council for PATT T\&S support and support from grant ST/J001082/1. RE is thankful to the NSF, Hungary (OTKA, Ref. No. K83133) and acknowledges M. K\'eray for patient encouragement. MM (QUB) acknowledges support from an STFC rolling grant and also would like to thank the Air Force Office of Scientific Research, Air Force Material Command, USAF for sponsorship under grant number FA8655-09-13085. MM (AO) thanks the Leverhulme Trust for support. Hinode is a Japanese mission developed and launched by ISAS/JAXA, with NAOJ as domestic partner and NASA and STFC (UK) as international partners. It is operated by these agencies in co-operation with ESA and NSC (Norway).

\end{document}